# Optimal Conditions for Observing Fractional Josephson Effect in Topological Josephson Junctions

Yeongmin Jang[1], Yong-Joo Doh[1*]

[1]Department of Physics and Photon Science, Gwangju Institute of Science and Technology (GIST), Gwangju 61005, Korea

**Corresponding Authors**

[*]E-mail: yjdoh@gist.ac.kr

# ABSTRACT

Topological Josephson junctions (JJs), which contain Majorana bound states, are expected to exhibit 4π-periodic current-phase relation, thereby resulting in doubled Shapiro steps under microwave irradiation. We performed numerical calculations of dynamical properties of topological JJs using a modified resistively and capacitively shunted junction model and extensively investigated the progressive evolution of Shapiro steps as a function of the junction parameters and microwave power and frequency. Our calculation results indicate that the suppression of odd-integer Shapiro steps, i.e., evidence of the fractional ac Josephson effect, is enhanced significantly by the increase in the junction capacitance and $I_cR_n$ product as well as the decrease in the microwave frequency even for the same portion of the 4π-periodic supercurrent. Our study provides the optimal conditions for observing the fractional ac Josephson effect; furthermore, our new model can be used to precisely quantify the topological supercurrent from the experimental data of topological JJs.

## I. INTRODUCTION

When a topological material is proximity-coupled to a conventional superconductor, topological superconductivity can be formed to host Majorana bound states (MBS) [1,2], which are expected to exhibit non-Abelian statistics and provide the basis for fault-tolerant quantum computation [3]. Topological Josephson junctions (JJs) [4], in which MBS are localized on either side of the junction, can carry supercurrents via single electrons instead of the typical Cooper pairs, resulting in supercurrents proportional to $\sin(\phi/2)$ rather than $\sin(\phi)$ [5], where $\phi$ denotes the gauge-invariant superconducting phase difference across the JJ. The $4\pi$-periodic current-phase relation (CPR) exhibits the fractional ac Josephson effect [5], in which the Josephson frequency $f_J$ is expressed as $eV/h$ instead of $2eV/h$, where $e$ is the elementary charge, $h$ the Planck constant, and $V$ the voltage difference across the JJ. Consequently, Shapiro steps due to the phase locking between the JJ and external microwave [6] are expected to be quantized in even integer multiples of $hf/2e$ for topological JJ [5].

Proximity-coupled JJs based on one-dimensional semiconductor nanowires [7,8], topological insulators [9,10], and Dirac semimetals [11,12] can provide a useful platform for realizing topological JJs. To date, doubled Shapiro steps due to topological superconductivity have been demonstrated using topological JJs made of InSb nanowires [13], HgTe-based topological insulators [14,15], and Dirac semimetals of $Bi_{1-x}Sb_x$ [16] and $Cd_3As_2$ [17]. Although the absence of odd-integer Shapiro steps is expected in theory under ideal conditions, the suppression of odd-integer steps and particularly the absence of the first Shapiro step, $V_1 = hf/2e$, were mainly observed in previous experiments [13-17]. Moreover, other JJs using similar topological materials and device geometry exhibited only conventional Shapiro steps without the first-step missing behavior

[18-21]. Hence, it is critical to study the optimal conditions for observing the fractional ac Josephson effect.

There have been several attempts to explain previous experimental observations based on the assumptions of the coexistence of $2\pi$- and $4\pi$-periodic supercurrents and the resistively-shunted junction (RSJ) model as an equivalent circuit of topological JJs [14,22]. Although the modified RSJ model successfully explained the experimental observations of the first-step missing behavior and its microwave frequency dependence, it failed to reflect the hysteretic current ($I$) vs. voltage ($V$) curves, which are frequently observed in topological JJs [13-15,23]. Because hysteretic dc $I$–$V$ curves indicate the existence of a capacitive coupling term in topological JJs [24,25], the RSJ model must be modified to include a capacitive term for a better understanding of topological JJs.

In this study, we extensively investigated the dynamical behavior of topological JJs under the irradiation of microwave using numerical calculations. We used a modified resistively and capacitively shunted junction (RCSJ) model that included a $4\pi$-periodic topological supercurrent and a junction capacitance, which contributed to hysteretic $I$–$V$ curves observed in previous experiments [24,25]. Our calculation results indicated that the suppression of odd-integer Shapiro steps owing to the topological Josephson effect improved significantly by the increase in the junction capacitance and $I_cR_n$ product, where $I_c$ denotes the critical current of the JJ, and $R_n$ the junction resistance in the normal state. Moreover, the odd-step suppression behavior was extremely sensitive to the microwave frequency even for the same portion of the topological supercurrent. Furthermore, our modified RCSJ model was used to quantitatively estimate the ratio of the $4\pi$-periodic supercurrent to the total supercurrent. It was revealed that the previous estimate based on the RSJ model was highly overestimated because capacitive coupling was disregarded in the JJ. Our results would be useful to design the optimal experimental conditions and to precisely estimate the topological supercurrent.

## II. THEORY

The conventional single JJ can be described by an equivalent circuit based on the RCSJ model, wherein a resistor, a capacitor, and the JJ itself as a channel for the supercurrent $I_{sc}$ are connected in parallel with each other [24]. In the presence of MBS, the supercurrent is expressed as the sum of two different types of supercurrents, i.e., $I_{sc} = I_{c,2\pi}\sin(\phi) + I_{c,4\pi}\sin(\phi/2)$, which corresponding to $2\pi$-periodic conventional and $4\pi$-periodic topological supercurrents, respectively [26]. Here, $I_{c,2\pi}$ and $I_{c,4\pi}$ are the critical currents of the $2\pi$- and $4\pi$-periodic supercurrents, respectively. Figure 1a shows the equivalent circuit of the topological JJ based on the two-supercurrent RCSJ model. By introducing new variables $I_c = I_{c,2\pi} + I_{c,4\pi}$ and the $4\pi$-periodic supercurrent parameter $\alpha_{4\pi} = I_{c,4\pi}/I_c$, the total supercurrent can be rewritten as $I_{sc} = (1 - \alpha_{4\pi})I_c\sin(\phi) + \alpha_{4\pi}I_c\sin(\phi/2)$.

When the JJ is biased with dc and ac currents, the equation of motion can be expressed as $I_{dc} + I_{ac}\sin(2\pi f t) = I_{sc} + CdV/dt + V/R_n$ according to Kirchhoff's law, where $f$ is the frequency of the ac current (or microwave), and $V$ is the voltage developed across the junction [24]. Using the ac Josephson relation $V = (\hbar/2e)\partial\phi/\partial t$, where $\hbar$ is the reduced Planck's constant, the equation of motion of the JJ can be expressed by

$$i_{sc}(\phi) + \frac{d\phi}{d\tau} + \beta_c\frac{d^2\phi}{d\tau^2} = i_{dc} + i_{ac}\sin(\Omega\tau), \qquad (1)$$

where we introduce a dimensionless current $i_k = I_k/I_c$ ($k = sc, dc, ac$), a dimensionless time $\tau = 2eI_cR_n t/\hbar$, the McCumber parameter $\beta_c = 2eI_cR_n^2 C/\hbar$, and a dimensionless frequency $\Omega = f/f_c = hf/2eI_cR_n$. Herein $f_c = 2eI_cR_n/h$ is the characteristic frequency of the JJ. In the present model, we assumed that the JJ was located at zero temperature to avoid any thermal noise current and thermal activation effects. In this study, we conducted numerical calculations of $\phi(t)$ based on Eq. (1) using the Runge–Kutta method and obtained $i_{dc}$ vs. $<V>$ characteristic curves as a function of $\alpha_{4\pi}$, $\beta_c$, and $\Omega$, where $<V>$ was obtained from the time-averaged Josephson relation $<V> = (\hbar/2e)<\partial\phi/\partial t>$.

### III. RESLTS AND DISCUSSION

Representative $i_{dc}$ vs. $<V>$ curves obtained from conventional RCSJ model ($\alpha_{4\pi} = 0$) for the cases of $\beta_c = 0.1, 1.0$, and $1.6$ are shown in Figure 1b. When the dc bias current was increased from $i_{dc} = 0$, $<V>$ jumped from the supercurrent branch to the dissipative one at switching current $i_s$ in all three cases. However, as $i_{dc}$ was decreased from the resistive branch, the $i_{dc}$ vs. $<V>$ curves with $\beta_c = 1.0$ and $1.6$ exhibited a hysteretic behavior, returning to the supercurrent branch at the return current $i_r$ smaller than $i_s$, whereas the $i_{dc}$ vs. $<V>$ curve with $\beta_c = 0.1$ was nonhysteretic. Figure 1c shows the relation between the parameter $\beta_c$ and hysteresis ratio $i_r/i_s$, confirming that our numerical calculations (symbols) of the RCSJ model agreed well with the approximate formula in Ref. [27]. Because the experimental values of the hysteresis parameter were reported to be $i_r/i_s = 0.6 – 1.0$ [13-15,17,23], the corresponding $\beta_c$ ranged from 0.9 to 3.8, as shown in Figure 1c.

Under microwave irradiation at frequency $f$, the $i_{dc}$ vs. $<V>$ curves are expected to exhibit Shapiro steps at voltage $<V> = nhf/2e$, where $n$ is an (even) integer for purely $2\pi$- ($4\pi$-) periodic CPRs [22]. Figure 1d shows our calculation results of Shapiro steps of nonhysteretic ($\beta_c = 0.1$) JJs for the cases of $\alpha_{4\pi}$ = 0, 0.3, and 1.0. It is evident that voltage plateaus with voltage interval $\Delta V$ = $hf/2e$ ($hf/e$) appeared in the 2π- (4π-) periodic JJ, corresponding to $\alpha_{4\pi}$ = 0 (1.0), as expected from previous studies [22,26]. Moreover, we observed that the first Shapiro step with $n$ = 1 was absent for the mixed case of $\alpha_{4\pi}$ = 0.3, similar to experimental observations as an evidence of the 4π-periodic CPR [13-16]. Because the absence of the first Shapiro step was affected by several parameters, such as $\alpha_{4\pi}$, $\beta_c$, $i_{ac}$, and $\Omega$, more extensive studies based on the two-supercurrent RCSJ model are required to optimize the experimental conditions.

With the 4π-periodic parameter $\alpha_{4\pi}$ increased, the progressive change in the Shapiro steps are shown as colored plots in Figures 2a–2d for the cases of $\Omega$ = 0.1 and $\beta_c$ = 0.1, in which different color means different voltage steps satisfying the ac Josephson relation $<V>$ = $nhf/2e$ with integer $n$. It is discernible that the odd-integer Shapiro steps were suppressed, whereas the even ones enhanced with $\alpha_{4\pi}$. The normalized current width of the $n$-th Shapiro step, $\Delta i_n = \Delta I_n/I_c$, is depicted as a function of the microwave amplitude $i_{ac}$ in Figures 2e–2h for $n$ = 0, 1, 2, and 3. For the purely 2π-periodic CPR with $\alpha_{4\pi}$ = 0, $\Delta i_n$ resembled the $n$-th order Bessel function [7,24], as shown in Figure 2e, in which the peaks of the even-integer steps coincided with the dips of the odd-integer steps, or vice versa. However, for the case of a purely 4π-periodic CPR with $\alpha_{4\pi}$ = 1, the odd-integer steps were absent and the neighboring even-integers steps were modulated in the out-of-phase mode with each other, i.e., the peaks (dips) of $\Delta i_0$ occurred concurrently with the dips (peaks) of $\Delta i_1$. For the mixed supercurrent case with 0 < $\alpha_{4\pi}$ < 1, the odd-integer steps were suppressed, whereas the even-integer steps enhanced. Additionally, a residual supercurrent $\Delta i_{0,res}$ appeared in the first lobe of $\Delta i_0$, as shown in Figure 2f, which was consistent with experimental observations [17,23] as an evidence of the 4π-periodic supercurrent.

We now turn to the case of hysteretic JJs. The numerical calculations of the Shapiro steps in the hysteretic JJs are depicted as colored plots in Figures 3a–3b for the case of $\beta_c = 1.0$. Compared with the calculation results obtained from the nonhysteretic JJs with the same $\alpha_{4\pi}$ in Figure 2, it is clearly shown that the odd-integer steps were significantly suppressed in the hysteretic JJs. Consequently, $\Delta i_n$ in the hysteretic JJ with $\alpha_{4\pi} = 0.15$ in Figure 3c appeared similar to $\Delta i_n$ in the nonhysteretic JJ with $\alpha_{4\pi} = 0.3$ in Figure 2g, indicating that the first-step missing behavior became more sensitive to $\alpha_{4\pi}$ in the hysteretic JJ with larger $\beta_c$. More details are discussed below.

As an indicator of the suppression of the odd-integer Shapiro steps, we introduced a parameter $\Delta i_{n,max}/\Delta i_{2,max}$ with $n = 1$ and 3, where $\Delta i_{n,max}$ is the maximum $\Delta i_n$ value in the first lobe of the $\Delta i_n$ vs. $i_{ac}$ curve. Figure 3e shows $\Delta i_{1,max}/\Delta i_{2,max}$ as a function of $\alpha_{4\pi}$ for different $\beta_c$ values. It is noteworthy that the first step was missing at $\alpha_{4\pi} = 0.5$, 0.2, and 0.15 for $\beta_c = 0.1$, 1.0, and 1.6, respectively, indicating that highly hysteretic JJs are more advantageous for observing the absence of the first Shapiro step even with the smaller fraction of the $4\pi$-periodic supercurrent. When we calculated $\Delta i_{n,max}/\Delta i_{2,max}$ ($n = 1, 3$) by increasing $\beta_c$ for a fixed $\alpha = 0.1$, it was clearly observed that the suppression of the odd-integer Shapiro steps enhanced with larger $\beta_c$, as depicted in Figure 3f. Additionally, the first Shapiro step was suppressed more sensitively to $\beta_c$ than the third one, suggesting the first-step suppression as a good indicator to optimize the measurement conditions for observing the fractional ac Josephson effect.

Because the McCumber parameter $\beta_c$ is proportional to $I_c R_n$ and $C$, the most dominant parameter for observing the first-step missing behavior must be identified. Thus we calculated the

Shapiro steps by increasing $\beta_c$ via two different methods: changing only $C$ while maintaining $I_cR_n$ product (see Figures 4a–4c), and vice versa (Figures 4a, 4d–4e). Other parameters, such as $\alpha_{4\pi} = 0.1$ and $f = 1$ GHz, were fixed at constant values for the calculations, whereas $I_cR_n$ product and $C$ were set based on the experimental values [15,16,23]. When we obtained $\Delta i_{1,max}/\Delta i_{2,max}$ as a function of $\beta_c$ for two different cases, the first Shapiro step was missing at $\beta_c = 0.7$ for the $I_cR_n$-varying case, which was much smaller than $\beta_c = 1.6$ obtained from the $C$-varying case (see Figure 4f). Hence, our calculation results suggest that forming a topological JJ with a larger $I_cR_n$ value is more efficient for observing fractional ac Josephson effect than adding a shunt capacitor to the JJ even for the same value of $\beta_c$. It is noteworthy that the increase in $I_cR_n$ product resulted in a decrease in the dimensionless frequency $\Omega = f/f_c$ in addition to an increase in $\beta_c$ because $f$ was fixed for the calculations. The dimensionless frequency was $\Omega = 0.1$, 0.05, and 0.03 for Figures 4a–4c, 4d, and 4e, respectively. A decreased $\Omega$ was also manifested by the narrower current widths $\Delta i_n$ in Figures 4d and 4e. Hence, we concluded that the $I_cR_n$ product was the crucial parameter for observing the first-step missing behavior in the topological JJs by affecting both $\beta_c$ and $\Omega$. The $\Omega$ dependence of the topological Shapiro steps will be discussed later.

The Shapiro steps in the topological JJ with $\alpha_{4\pi} = 0.1$ and $\beta_c = 1.0$ are shown in Figures 5a and 5b for the cases of $\Omega = 0.1$ and 0.4, respectively. The odd-integer Shapiro steps were highly suppressed under the microwave irradiation with $\Omega = 0.1$ in Figure 5a, but they recovered fully and became comparable to the even-integer steps with $\Omega = 0.4$ in Figure 5b. More detailed results are shown in Figure 5c, exhibiting $\Delta i_{n,max}/\Delta i_{2,max}$ ($n = 1$ and 3) as a function of $\Omega$ at fixed junction parameters $\alpha_{4\pi} = 0.1$ and $\beta_c = 1.0$. The first and third steps were missing at $\Omega = 0.08$ and 0.04, respectively. When $\Omega$ was increased up to $\Omega = 0.3$, the maximum Shapiro-step width $\Delta i_{n,max}$ exhibited similar values for all integer steps except $n = 0$, characteristic of conventional Shapiro steps, as shown in Figure 2a. Our calculation results indicate that the low-frequency microwave irradiation was also critical for observing the suppression of odd-integer Shapiro steps in the topological JJ, including the mixed supercurrents. This frequency dependence has already been reported in previous experiments [14-16,26]; it was attributed to the reduced characteristic frequency of the topological JJ, corresponding to $f_{c,4\pi} = 2eI_{c,4\pi}R_n/h = \alpha_{4\pi}f_c$. Theoretical study [28] also suggested that the suppression of the odd-integer Shapiro steps would be enhanced under the condition of $\Omega \leq \alpha_{4\pi}$, which is consistent with our calculation results shown in Figure 5c.

Our modified RCSJ model can be used to quantitatively estimate the $4\pi$-supercurrent parameter $\alpha_{4\pi}$ from the experimental data. In Figure 5d, we show the calculation results of $\Delta i_{n,max}/\Delta i_{2,max}$ ($n = 1$ and $3$) as a function of $\alpha_{4\pi}$ for a comparison with experimental values obtained in a previous study using HgTe-based topological JJs [14]. The fitting parameters $\beta_c = 1.76$ and $\Omega = 0.056$ were obtained using the experimental values of $i_r/i_s$, $I_c$, $R_n$, and $f$ from the same reference. From the experimental results of $\Delta i_{1,max}/\Delta i_{2,max} = 0$ and $\Delta i_{3,max}/\Delta i_{2,max} = 0.92$ in Ref. [14], we estimated that $\alpha_{4\pi} = 0.013$, i.e., the $4\pi$-periodic supercurrent was approximately 1.3% of $I_c$, which was much smaller than the previous estimate of ~ 9% obtained from the nonhysteretic RSJ model [14]. The difference between the two estimates is attributed to the enhanced sensitivity of the modified RCSJ model for detecting the topological Josephson effect. Additionally, we discovered that the previous RSJ model tended to overestimate the $4\pi$-periodic supercurrent ratio of the hysteretic JJs.

### IV. CONCLUSIONS

We investigated the dynamical properties of a capacitively coupled topological JJ using the

RCSJ model modified with the $4\pi$-periodic supercurrent and sought for the optimal conditions for observing the fractional ac Josephson effect. Our results suggested that the increase in the McCumber parameter $\beta_c$, which resulted in a hysteretic $I$–$V$ curve, enhanced the suppression of odd-integer Shapiro steps and the sensitivity for detecting the topological supercurrent parameter $\alpha_{4\pi}$. In particular, increasing $I_cR_n$ product rather than the junction capacitance $C$ was more efficient for observing the suppression of odd-integer steps for the same $\alpha_{4\pi}$. For the topological JJ with the mixed $2\pi$- and $4\pi$-periodic supercurrents, the optimal microwave frequency for observing the fractional ac Josephson effect decreased according to $\alpha_{4\pi}$. Our modified RCSJ model can be used to precisely estimate $\alpha_{4\pi}$ from the experimental data. We believe that our results would significantly benefit the investigation of the fractional ac Josephson effect in topological JJs.

## Acknowledgments

We are grateful to Ryundon Kim, Youngmin Cho, Sang-Jun Choi, and H.-S. Sim for the useful discussions. This research was supported by the NRF of Korea through the Basic Science Research Program (2018R1A3B1052827).

# FIGURE CAPTIONS

**Figure 1.** (a) Equivalent circuit of topological JJ in modified RCSJ model. $R_n$ is the junction resistance in the normal state, $C$ the junction capacitance, and $I_{SC,2\pi}$ ($I_{SC,4\pi}$) the $2\pi$ ($4\pi$)-periodic supercurrent. (b) Calculation results of current *vs.* voltage curves for McCumber parameter $\beta_c$ = 0.1, 1.0, and 1.6. Arrows indicate switching current ($i_s$) and return current ($i_r$). Red and blue curves were offset horizontally for clarity. (c) $\beta_c$ as a function of $i_r/i_s$, obtained from numerical calculations (symbols) using RCSJ model and approximate solutions in Ref. [27]. (d) Calculation results of current *vs.* voltage curves with Shapiro steps for $4\pi$-periodic supercurrent parameter $\alpha_{4\pi}$ = 0, 0.3, and 1.0. Other parameters were fixed at $\beta_c$ = 0.1, $\Omega$ = 0.1, and $i_{ac}$ = 1.0. Red and blue curves were offset horizontally for clarity.

**Figure 2.** (a–d) Color plots of voltage of two-supercurrent JJ as a function of $i_{dc}$ and $i_{ac}$ for $\alpha_{4\pi}$ = 0, 0.15, 0.3, and 1.0, for the nonhysteretic JJ with $\beta_c$ = 0.1. Dimensionless frequency was $\Omega$ = 0.1. Integer $n$ indicates order of Shapiro steps. (e–h) Current widths $\Delta i_n$ ($n$ = 0, 1, 2, and 3) of Shapiro steps shown in (a–d). Curves were offset vertically for clarity. $\Delta i_{n,max}$ is the maximum current width in the first lobe of $\Delta i_n$ ($n$ = 1, 2, and 3), and $\Delta i_{0,res}$ is the residual critical current in the first lobe of $\Delta i_0$.

**Figure 3.** (a–b) Color plots of topological JJ voltage as a function of $i_{dc}$ and $i_{ac}$ for $\alpha_{4\pi}$ = 0.15 and 0.3, for the hysteretic JJ with $\beta_c$ = 1.0. Dimensionless frequency was $\Omega$ = 0.1. Integer $n$ indicates order of Shapiro steps. (c–d) Current widths $\Delta i_n$ ($n$ = 0, 1, 2, and 3) of Shapiro steps, shown in (a–b). Curves were offset vertically for clarity. (e) Step-width ratio $\Delta i_{1,max}/\Delta i_{2,max}$ as a function of $\alpha_{4\pi}$ for different $\beta_c$ values. Solid lines provide visual guidance. (f) Step-width ratios $\Delta i_{1,max}/\Delta i_{2,max}$ (red

square) and $\Delta i_{3,\text{max}}/\Delta i_{2,\text{max}}$ (blue circle) based on $\beta_c$.

**Figure 4.** (a–e) Color plots of topological JJ voltage as a function of $i_{\text{dc}}$ and $i_{\text{ac}}$ with varying $C$ and $I_cR_n$. Other parameters were fixed at $\alpha_{4\pi} = 0.1$ and $f = 1$ GHz. (f) Step-width ratio $\Delta i_{1,\text{max}}/\Delta i_{2,\text{max}}$ as a function of $\beta_c$. $\beta_c$ was varied with increasing $C$ (blue circle) for fixed $I_cR_n = 20$ μV or increasing $I_cR_n$ (red square) for fixed $C = 82$ fF.

**Figure 5.** Current widths $\Delta i_n$ ($n = 0$, 1, 2, and 3) of Shapiro steps for (a) $\Omega = 0.1$ and (b) 0.4. Other parameters were fixed at $\alpha_{4\pi} = 0.1$ and $\beta_c = 1.0$. Curves were offset vertically for clarity. (c) Step-width ratios $\Delta i_{1,\text{max}}/\Delta i_{2,\text{max}}$ (red square) and $\Delta i_{3,\text{max}}/\Delta i_{2,\text{max}}$ (blue circle) based on $\Omega$ with fixed $\alpha_{4\pi} = 0.1$ and $\beta_c = 1.0$. (d) $\Delta i_{1,\text{max}}/\Delta i_{2,\text{max}}$ (red square) and $\Delta i_{3,\text{max}}/\Delta i_{2,\text{max}}$ (blue circle) as a function of $\alpha_{4\pi}$ compared with experimental data obtained from Ref. [14]. Other parameters were fixed at $\beta_c = 1.76$ and $\Omega = 0.056$, which were obtained from the same reference.

**Fig.1**

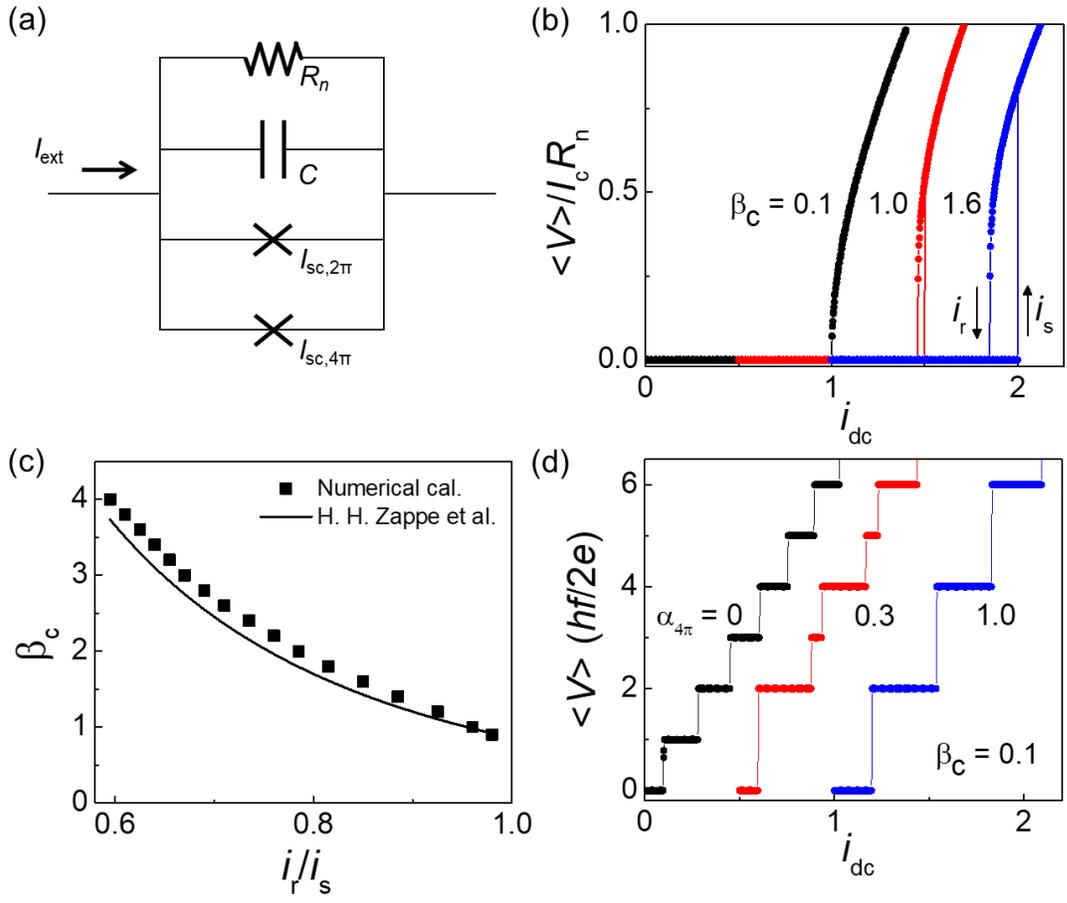

**Fig.2**

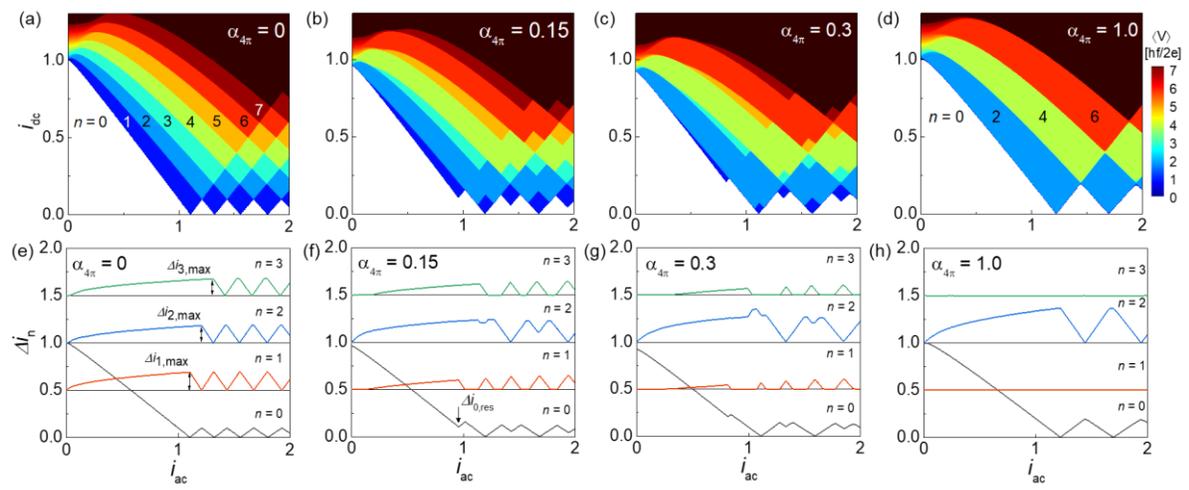

**Fig.3**

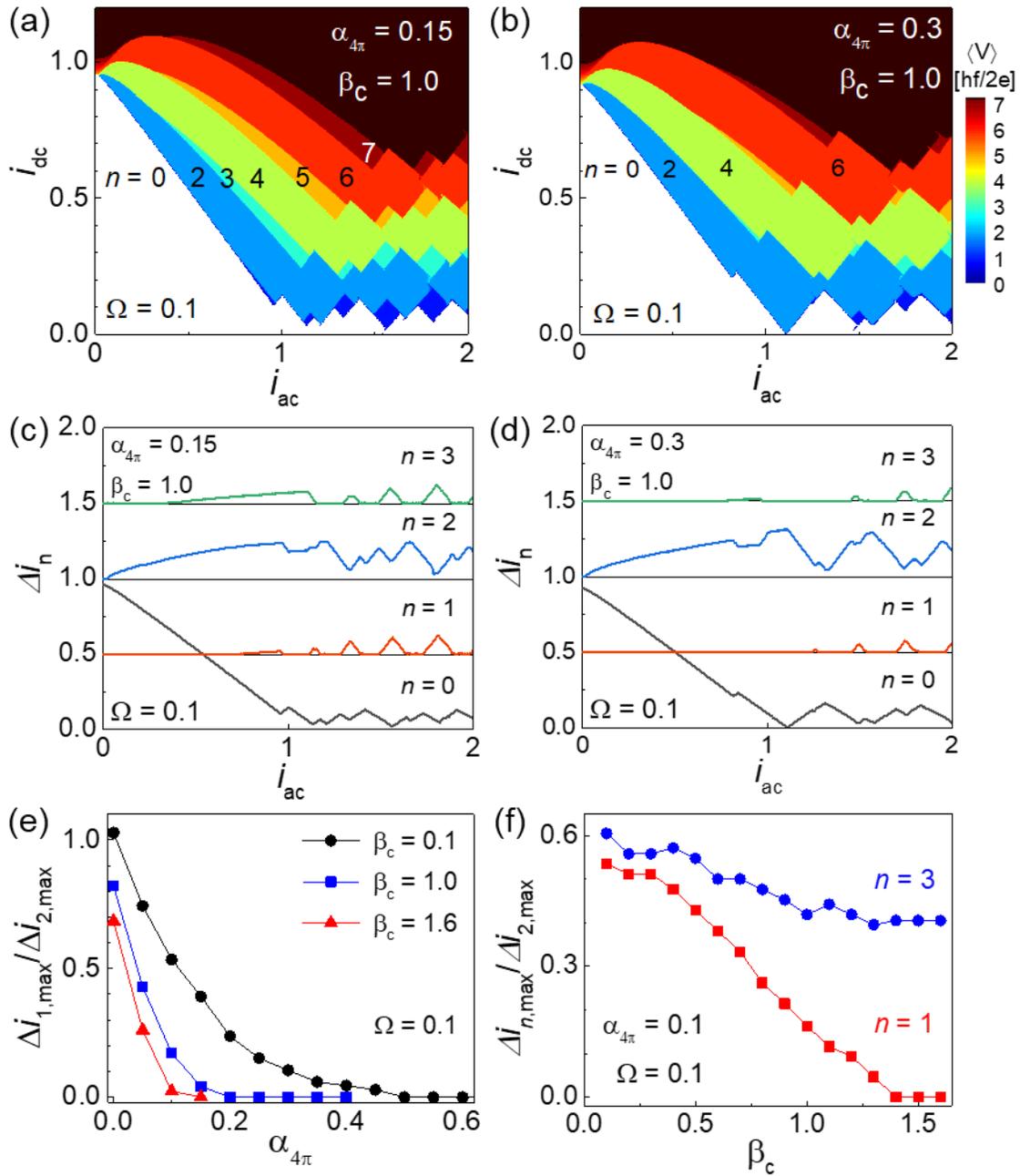

**Fig.4**

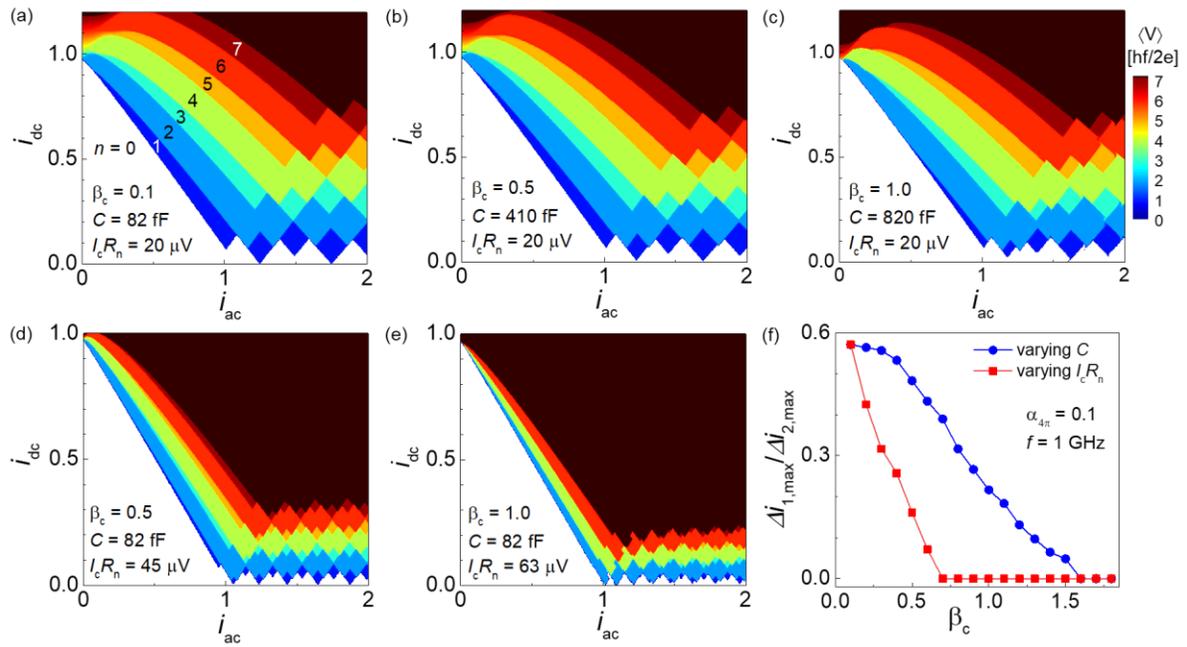

**Fig.5**

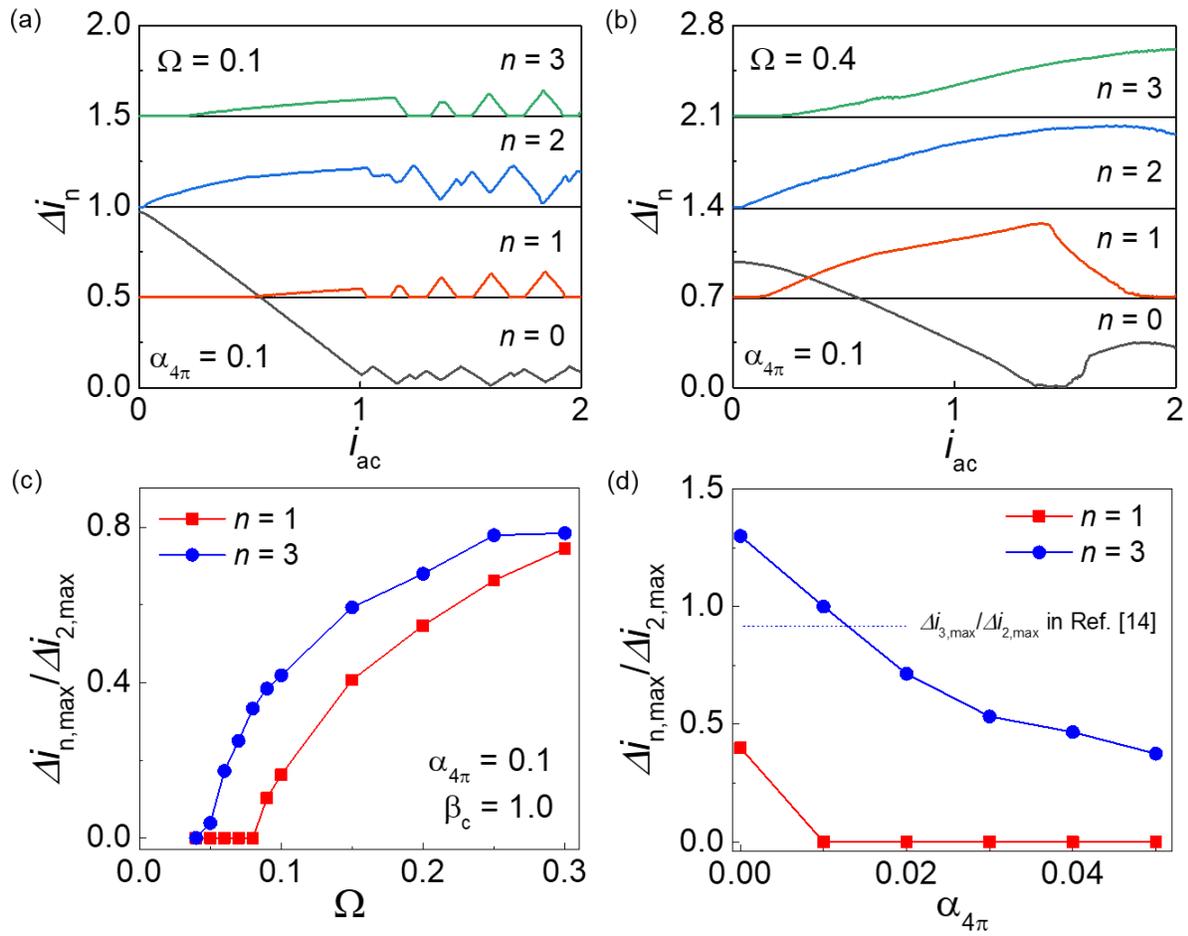